\documentclass{webofc}
\usepackage[varg]{txfonts}   % Web of Conferences font
\usepackage{hyperref}
\usepackage{url,xcolor}
\hypersetup{colorlinks=true,citecolor=blue,urlcolor=blue,linkcolor=blue}
\definecolor{MyBlue}{HTML}{0000FE}
\definecolor{MyGrey}{HTML}{999999}
\definecolor{MyOrange}{HTML}{FF9900}
\definecolor{MyDarkRed}{HTML}{CC0033}
\definecolor{MyPurple}{HTML}{A78EC2}
\definecolor{MyDarkGreen}{HTML}{006766}
\definecolor{MyCyan}{HTML}{00D1D1}
\definecolor{MyLightBlue}{HTML}{5785FF}
\definecolor{MyPink}{HTML}{FF00FE}

\providecommand{\abs}[1]{\lvert#1\rvert}    %        |.|

\begin{document}
\title{Theory on CKM and heavy quark decay}

\author{\firstname{Oliver} \lastname{Witzel}\inst{1}\fnsep\thanks{\email{oliver.witzel@uni-siegen.de}}}

\institute{Center for Particle Physics Siegen, Theoretische Physik 1,
Naturwissenschaftlich-Technische Fakultät, Universität Siegen, 57068 Siegen, Germany}

\abstract{The combination of precise experimental measurements and theoretical predictions allows to extract Cabibbo-Kobayashi-Maskawa (CKM) matrix elements or constrain flavor changing processes in the standard model. Focusing at theoretical predictions, we review recent highlights from the sector of heavy charm and bottom quark decays. Special emphasis is given to nonperturbative contributions due to the strong force calculated using lattice QCD.}
\maketitle
\section{Introduction}
\label{intro}
In the standard model (SM) of elementary particle physics quark masses and mixing arises from the Yukawa interactions with the Higgs condensate. The probability for the transition of a quark flavor $j$ to a flavor $i$ is encoded in the Cabibbo-Kobayashi-Maskawa (CKM) matrix. In the SM with three generations of quark flavors, the CKM matrix is a unitary $3\times 3$ matrix. Its elements are fundamental parameters of the SM which are determined combining experimental measurements and theoretical calculations. The following values refer to the 2022 review of particle physics by the Particle Data Group (PDG) \cite{ParticleDataGroup:2022pth}\footnote{The 2024 review has become available at \cite{ParticleDataGroup:2024cfk}.} 

\begin{align}
  \begin{bmatrix}
    V_{ud} & V_{us} & V_{ub} \\
    V_{cd} & V_{cs} & V_{cb} \\
    V_{td} & V_{ts} & V_{tb} 
  \end{bmatrix} 
  &\;=
  \begin{bmatrix}
    0.97370(14)&0.2245(8) &0.00382(24)\\
    0.221(4)   &0.987(11) &0.0408(14)\\
    0.0080(3)  &0.0388(11)&1.013(30)
  \end{bmatrix}.
\end{align}
While the most precisely known matrix element $|V_{ud}|$ is has better than per mille level precision, the least precisely known matrix element $|V_{ub}|$ is quoted with an uncertainty of $6.3\%$
\begin{align}            
  \frac{\abs{\delta V_{CKM}}}{\abs{V_{CKM}}} &=
  \begin{bmatrix}
    0.014&0.35&6.3\\
    1.8&1.1&3.4\\
    3.8& 2.8 & 3.0
  \end{bmatrix}\%.
\end{align}

\begin{figure}[t]
  \centering
    \begin{picture}(75,30)
      \put(3,3){\includegraphics[scale=0.5]{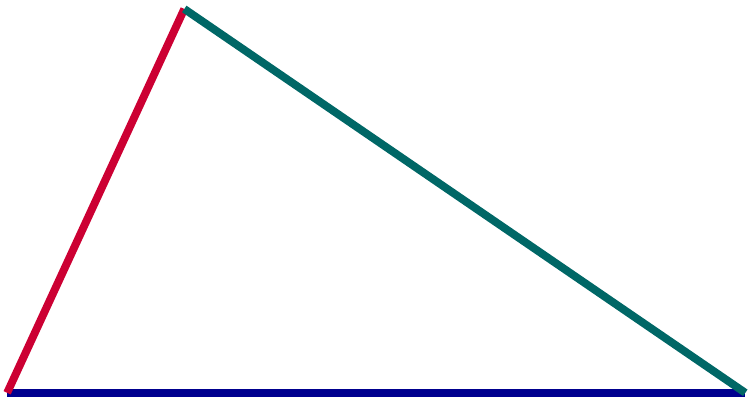}}
      \put(0,0){$(0,0)$}
      \put(63,0){$(1,0)$}
      \put(15,37.5){$(\bar\rho,\bar \eta)$}
      \put(16.5,29){$\alpha=\phi_2$}
      \put(5.5,5){$\gamma=\phi_3$}
      \put(52.5,5){$\beta=\phi_1$}      
      \put(1,24){\textcolor{MyDarkRed}{\large $\frac{\left|V_{ud}^{\phantom{*}}V_{ub}^*\right|}{\left|V_{cd}^{\phantom{*}}V_{cb}^*\right|}$}}
      \put(41,24){\textcolor{MyDarkGreen}{\large $\frac{\left|V_{td}^{\phantom{*}}V_{tb}^*\right|}{\left|V_{cd}^{\phantom{*}}V_{cb}^*\right|}$}}
    \end{picture}
  \caption{Sketch of the unitarity triange defined by Eq.~\eqref{Eq.UT}.}
  \label{Fig.UT}
\end{figure}
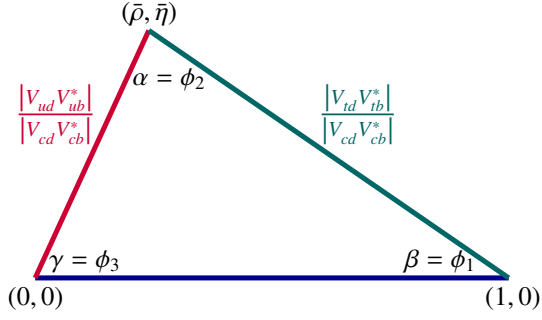

Following the discussion in \cite{ParticleDataGroup:2022pth}, we can exploit the fact that the CKM matrix in the SM is unitary and parametrize it in different ways. A popular choice expresses the CKM matrix in terms of three mixing angles and a CP-violating phase  
\begin{align}
  V_\text{CKM} = \begin{bmatrix}
    c_{12} c_{13} & s_{12}c_{13} & s_{13}e^{-i\delta} \\
    -s_{12}c_{23} - c_{12}s_{23}s_{13} e^{i\delta} & c_{12}c_{23}-s_{12}s_{23}s_{13}e^{i\delta}& s_{23}c_{13} \\
    s_{12}s_{23} - c_{12}c_{23}s_{13} e^{i\delta} & -c_{12}s_{23}-s_{12}c_{23}s_{13}e^{i\delta}& c_{23}c_{13}
  \end{bmatrix}.
  \label{Eq.CKM_angels_phase}
\end{align}
If we acknowledge the experimental observation that
\begin{align}
  s_{13} \ll s_{23} \ll s_{12} \ll 1,
\end{align}
we can highlight the hierarchical nature of the CKM matrix and arrive at the \emph{Wolfenstein parametrization}\/
\begin{align}
  V_\text{CKM} = \begin{bmatrix}
    1-\lambda^2/2 & \lambda & A\lambda^3(\rho -i\eta) \\
    -\lambda & 1-\lambda^2/2 & A\lambda^2 \\
    A\lambda^3(1-\rho -i\eta) & -A\lambda^2 &1 
  \end{bmatrix} + {\cal O}(\lambda^4),
  \label{Eq.Wolfenstein}
\end{align}
which is unitary in all order of $\lambda$. In Eqs.~\eqref{Eq.CKM_angels_phase} -- \eqref{Eq.Wolfenstein} we used the following notation: 
\begin{align}
  s_{12} &= \lambda = \frac{|V_{us}|}{|V_{ud}|^2+|V_{us}|^2}, \qquad%\notag \\
  s_{23} =A\lambda^2 = \lambda \left| \frac{V_{cb}}{V_{us}}\right|, \\
  s_{13} e^{i\delta} &= V^*_{ub} = A\lambda^3 (\rho + i\eta) = \frac{A\lambda^3(\bar \rho + i\bar\eta)\sqrt{1-A^2\lambda^4}}{\sqrt{1-\lambda^2}\left(1-A^2\lambda^4(\bar\rho + i\bar\eta)\right)},\qquad%\notag \\
  (\bar\rho +i\bar\eta) &= -\frac{V_{ud}V_{ub}^*}{V_{cd}^{\phantom{*}}V_{cb}^*}. \notag
\end{align}
The virtue of this form is to visualize the unitary CKM matrix in terms of six different unitarity triangles. Most commonly used is the one based on the relation
\begin{align}
  V_{ud}^{\phantom{*}}V_{ub}^* + V_{cd}^{\phantom{*}}V_{cb}^* + V_{td}^{\phantom{*}}V_{tb}^* =0.
  \label{Eq.UT}
\end{align}
Dividing all sides by $V_{cd}V_{cb}^*$, the vertices are exactly at $(0,0)$, $(1,0)$, and $(\bar \rho,\bar\eta)$ as shown in the sketch in Fig.~\ref{Fig.UT}. The quest is now to over-constrain CKM elements in order to test and constrain the SM. Two groups, CKMfitter \cite{Charles:2004jd,CKMfitter} and UTfit \cite{UTfit:2022hsi,UTfit}, regularly gather experimental and theoretical updates to perform global fits of the CKM unitarity triangle. 

In the following sections we discuss updates on the determinations of the CKM matrix elements $|V_{cd}|$, $|V_{cb}|$, and $|V_{ub}|$ which all involve either a heavy charm or bottom quark before summarizing in Section \ref{summary}.

\section{Determination of \texorpdfstring{$V_{cd}$}{Vcd}}
\label{Vcd}
First we consider the determination of $V_{cd}$ for which the PDG \cite{ParticleDataGroup:2022pth} presently reports an uncertainty of 1.8\%. The PDG averages three different determinations:
\begin{itemize}
\item Determinations based on neutrino scattering data: $|V_{cd}|_{PDG}^{\nu} = 0.230 \pm 0.011$
\item Leptonic $D^+\to \{\mu^+ \nu_\mu, \tau^+ \nu_\tau\}$ decays: $|V_{cd}|_{PDG}^{f_D} = 0.2181 \pm 0.0050$
\item Semileptonic $D\to \pi\ell\nu$ decays (at $q^2=0$): $|V_{cd}|_{PDG}^{D\pi(0)} = 0.233 \pm 0.014$ 
\end{itemize}
which results at the value of
\begin{align}
  |V_{cd}|_{PDG} =  0.221 \pm 0.004.
\end{align}

The determinations based on leptonic (semileptonic) decays are obtained by combining experimental data and theoretical calculations of decay constants (form factors), using lattice quantum chromodynamics (LQCD). In the case of leptonic decays, experimental data from BESIII \cite{BESIII:2013iro} and CLEO \cite{CLEO:2008ffk} are combined with LQCD calculations by Fermilab/MILC \cite{Bazavov:2017lyh} and ETMC \cite{Carrasco:2014poa}. For semileptonic decays measurements by BaBar \cite{BaBar:2014xzf}, BESIII \cite{BESIII:2015tql,BESIII:2017ylw}, CLEO-c \cite{CLEO:2009svp}, and Belle \cite{Belle:2006idb} as well as the LQCD form factors by ETMC \cite{Lubicz:2017syv} at $q^2=0$ are used.

Recently the Fermilab/MILC collaboration published new results determining the form factors  $D\to \pi\ell\nu$ and $D_s\to K\ell\nu$ over the full $q^2$ range \cite{FermilabLattice:2022gku}. Combining the $D\to\pi\ell\nu$ form factors with the experimental data from BaBar, BESIII, CLEO-c, and Belle \cite{BaBar:2014xzf,BESIII:2015tql,BESIII:2017ylw,CLEO:2009svp,Belle:2006idb} leads to a new most precise determination of
\begin{align}
  |V_{cd}|_{FNAL/MILC}^{D\pi} = 0.2238 \pm 0.0029.
\end{align}
The gain in precision arises by exploiting the full $q^2$ dependence in combination with state-of-the-art lattice simulations.\footnote{A possible point of concern is using $f_\parallel$ and $f_\perp$ in the chiral-continuum extrapolation (cf.~discussion in Sec.~\ref{Vub}).} In addition a first prediction of $|V_{cd}|$ based on semileptonic $D_s$ decays is presented and a value of 
\begin{align}
  |V_{cd}|_{FNAL/MILC}^{D_s K} = 0.258 \pm 0.015,
\end{align}
is obtained using experimental results by BESIII \cite{BESIII:2018xre}. Due to fewer experimental results with larger uncertainty, the precision of this channel is however limited.

\begin{figure}[t]
  \centering
  \includegraphics[height=0.15\textheight]{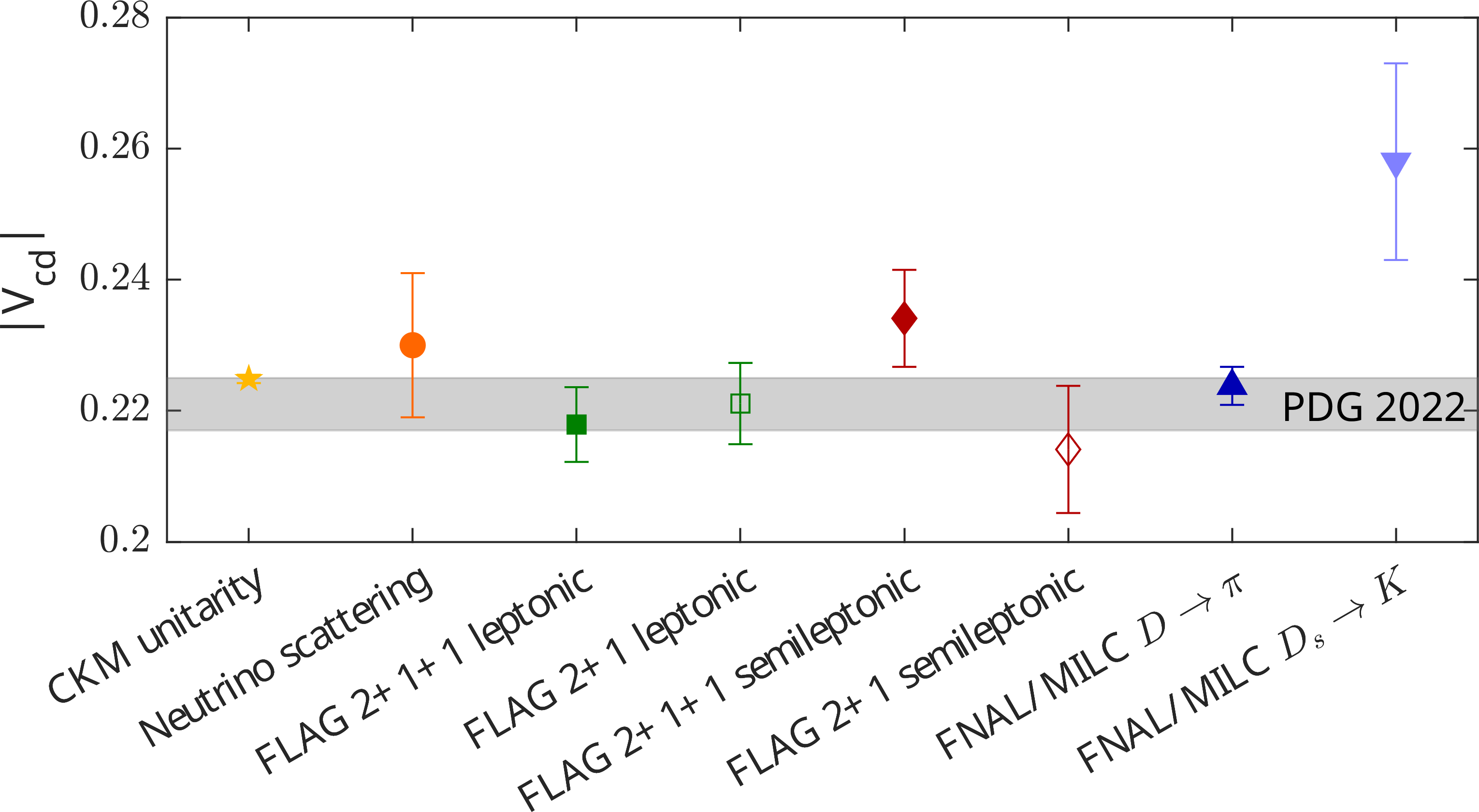}    
  \caption{Comparison of different $|V_{cd}|$ determinations. The value for `CKM unitarity' and `neutrino scattering' are taken from \cite{ParticleDataGroup:2022pth}, leptonic and semileptonic values refer to FLAG averages \cite{FlavourLatticeAveragingGroupFLAG:2021npn} for 2+1+1 \cite{Carrasco:2014poa,Bazavov:2017lyh,Lubicz:2017syv,Riggio:2017zwh,Chakraborty:2021qav} and 2+1 \cite{FermilabLattice:2011njy,Davies:2010ip,Na:2012iu,Yang:2014sea,Boyle:2017jwu,Na:2011mc,Na:2010uf}  flavors, respectively. The new, most precise determination of $|V_{cd}|$ based on $D\to\pi\ell\nu$ decays by Fermilab/MILC \cite{FermilabLattice:2022gku} is in excellent agreement with previous results.}
  \label{Fig.Vcd}
\end{figure}

Overall the different determinations of $|V_{cd}|$ show very good agreement as can be seen in the comparison plot shown in Fig.~\ref{Fig.Vcd}.

\section{Determination of \texorpdfstring{$V_{cb}$}{Vcb}}
\label{Vcb}

\begin{figure}[t]
    \centering
    \includegraphics[height=0.15\textheight]{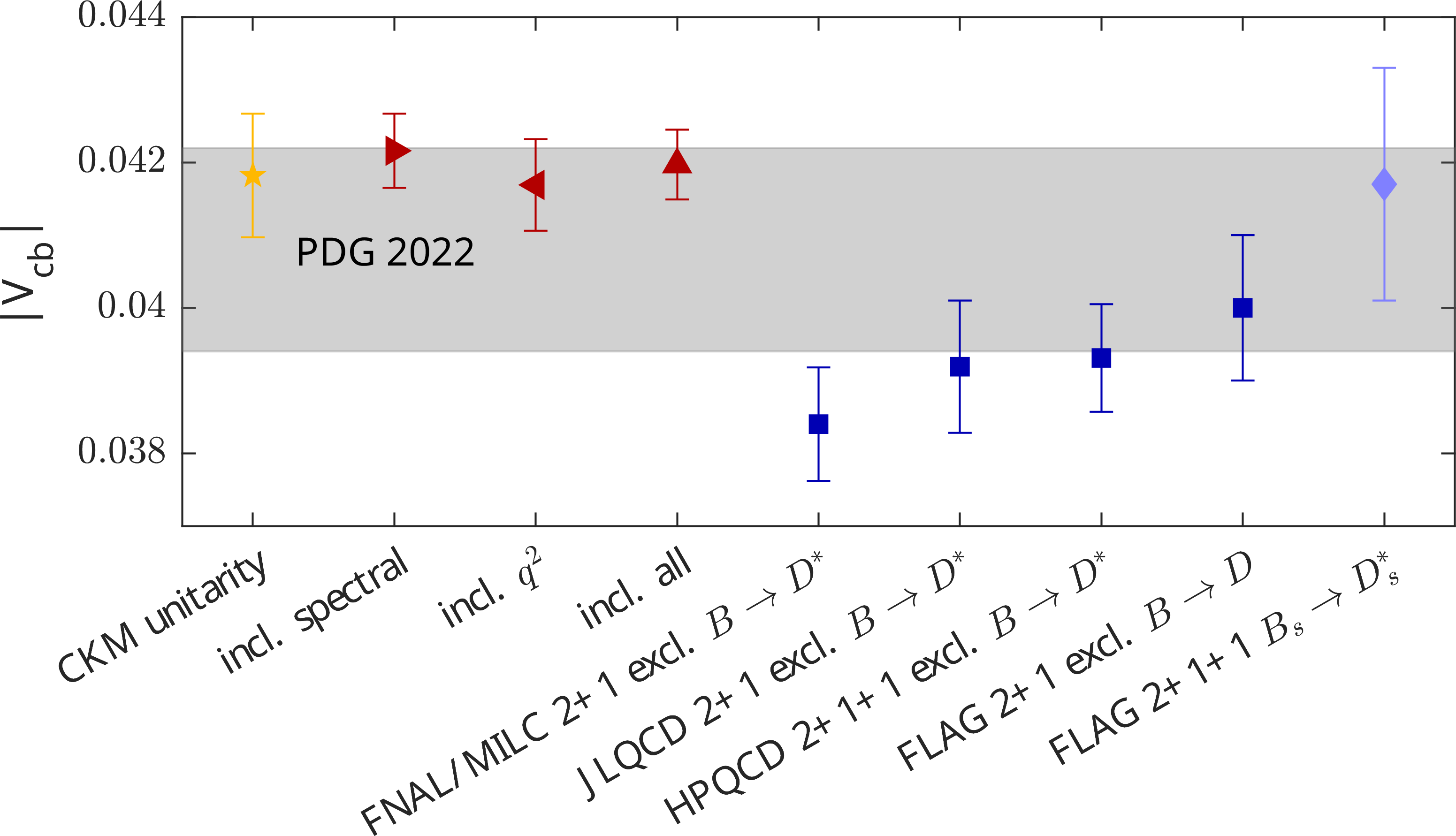}
    \caption{Compilation highlighting the tension between the inclusive determinations (red triangles) \cite{Bordone:2021oof,Bernlochner:2022ucr,Finauri:2023kte} and different exclusive channels (blue squares) to obtain $|V_{cb}|$. Shown are $B\to D^*$ determinations \cite{FermilabLattice:2021cdg,Aoki:2023qpa,Harrison:2023dzh} as well as FLAG averages for $B\to D$ and $B_s\to D_s^{(*)}$ \cite{FlavourLatticeAveragingGroupFLAG:2021npn,FLAG2024,FermilabLattice:2014ysv,Na:2015kha,MILC:2015uhg,McLean:2019sds,McLean:2019qcx}.}
    \label{Fig.Vcb}
\end{figure}

Unlike for $|V_{cd}|$, we cannot determine $|V_{cb}|$ from simple leptonic decays because an experimental measurement of $B_c \to \tau \nu_\tau$ is currently not feasible. Determinations of $|V_{cb}|$ are, therefore, based on analyzing semileptonic decays and we can consider both, inclusive and exclusive, processes. While in the case of exclusive decays the hadronic final state is explicitly specified, inclusive decays consider all semileptonic decays 
featuring a $b\to c$ transition. Unfortunately, the value obtained for $|V_{cb}|$ based on inclusive analyses has been showing a persistent $2-3\sigma$ tension to values corresponding to exclusive analyses. The current situations is summarized in Fig.~\ref{Fig.Vcb} where we show the values of inclusive determinations discussed below as well as FLAG averages \cite{FlavourLatticeAveragingGroupFLAG:2021npn,FLAG2024} for different exclusive channels.

\subsection{Inclusive determination of \texorpdfstring{$|V_{cb}|$}{Vcb}}
Measurements of inclusive $B \to X_c \ell\nu_\ell$ decays are typically performed at $B$-factories where an $e^+$ beam collides with an $e^-$ beam and the collision energy is tuned to the $\Upsilon(4s)$ threshold. The  $\Upsilon(4s)$ predominantly decays into $B$ and $\overline B$ mesons and their semileptonic decays are then experimentally observed. For the inclusive determination of $|V_{cb}|$ moments e.g.~of the out-going leptons are experimentally measured. $|V_{cb}^\text{incl}|$ is then extracted by fitting these lepton moments using a fit ansatz based on the systematic expansion of the total decay rate. This operator product expansion (OPE) is performed in terms of $\Lambda_\text{QCD}/m_b$ with $m_b \gg \Lambda_{\text{QCD}}$ and therefore named heavy quark expansion (HQE)
\begin{align}
  {\cal B} = |V_{cb}|^2 \left[ \Gamma(b\to c\ell \nu_\ell) + \frac{1}{m_{c,b}} + \alpha_s + \ldots\right].
  \label{Eq.HQE}
\end{align}
As is the case for all OPE, Eq.~\eqref{Eq.HQE} does not allow point-by-point predictions. It however converges if integrated over large phase space
\begin{align}
  \int d\Phi\; w^n(\nu, p_\ell, p_\nu) \frac{d\Gamma}{d\Phi} \quad \text{with} \quad \nu = p_B/m_B.
  \label{Eq.HQE_int}
\end{align}
In Eq.~\eqref{Eq.HQE_int} we have introduced a weight functions $w$ which can e.g.~be defined by
\begin{itemize}
\item 4-momentum transfer squared:  $w = (p_\ell + p_\nu)^2 = q^2$,
\item Invariant mass squared: $w = (m_B\nu -q)^2 = M_X^2$,
\item Lepton energy: $w = (\nu \cdot p_\ell) = E_\ell^B$. 
\end{itemize}

This method has been established using spectral moments (hadronic mass moments, lepton energy moments, \ldots)
\begin{align}
  d\Gamma = d\Gamma_0 + d\Gamma_{\mu_\pi} \frac{\mu^2_\pi}{m_b^2}
  + d\Gamma_{\mu_G} \frac{\rho^3_D}{m_b^3}
  + d\Gamma_{\rho_{LS}} \frac{\rho^3_{LS}}{m_b^3} + {\cal O}\left(\frac{1}{m_b^4}\right).
  \label{Eq.dGamma}
\end{align}
In Eq.~\eqref{Eq.dGamma} the $d\Gamma$ have been calculated perturbatively up to ${\cal O}(\alpha_s^3)$ \cite{Fael:2020tow} whereas $\mu_\pi^2$, $\mu_G^2$, $\rho_D^3$, $\rho_{LS}^3$ parameterize nonperturbative dynamics which is fitted from data. The state-of-the-art analysis including 3-loop $\alpha_s$ corrections for the semileptonic fit to experimentally measured spectral moments yields \cite{Bordone:2021oof}
\begin{align}
  |V_{cb}^\text{incl}| = (42.16 \pm 0.51) \cdot 10^{-3},
\end{align}
which has an uncertainty 1.2\%. Due to the large number of higher order terms in the HQE expansion it is, however, not straight-forward to further improve this determination.

The number of terms can be reduced by using reparametrization invariance (RPI) as proposed by Fael, Mannel, and Vos in Ref.~\cite{Fael:2018vsp}. Unfortunately, not all observables are RPI invariant. Out of the three weight functions named above, only the $q^2$ moments are RPI invariant. By now Belle \cite{Belle:2021idw} and Belle II \cite{Belle-II:2022evt} have performed dedicated analyses extracting the $\langle (q^2)^n\rangle$ moments and thus enabled the first determination of $|V_{cb}|$ using $q^2$ moments \cite{Bernlochner:2022ucr}. Including contributions up to $1/m_b^4$ and correction up to $\alpha_s$
\begin{align}
  |V_{cb}^\text{incl, $q^2$}| = (41.69 \pm 0.63) \cdot 10^{-3},
\end{align}
is obtained which has a competitive uncertainty of 1.5\%.

Simultaneously extracting $|V_{cb}^\text{incl}|$ using all moments, an even more precise value can be obtained \cite{Finauri:2023kte}
\begin{align}
  |V_{cb}^\text{incl, all}| = (41.97 \pm 0.48) \cdot 10^{-3},
\end{align}
which has an uncertainty of 1.1\%. We emphasize that the new determination based on $q^2$ moments provides a different lever arm to constrain the fit parameters than the method based on spectral moments.

\subsection{Exclusive determination of \texorpdfstring{$|V_{cb}|$}{Vcb}}
Exclusive decays have been measured experimentally both, at $B$ factories as well has at hadron colliders e.g.~the LHCb experiment at the large hadron collider (LHC). Such measurements have been reported with $B$, $B_s$, or $\Lambda_b$ initial states and pseudoscalar or vector hadronic final states. To extract $|V_{cb}^\text{excl}|$, these measurements need to be combined with form factors either determined using LQCD or determinations based on sum rules. In the following we restrict ourselves to exclusive $B \to D^{*} \ell\nu_\ell$ decays where the $D^*$ is treated as a QCD-stable particle using the narrow width approximation and form factors are obtained using LQCD. Experimentally $B\to D^* \ell \nu$ is preferred and measurements have been reported by BaBar, Belle, and Belle II \cite{BaBar:2019vpl,Belle:2018ezy,Belle-II:2023okj}.

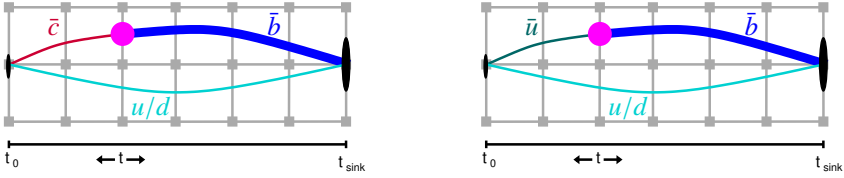
\begin{figure}[t]
  \centering
  \begin{picture}(110,25)
    \put(0,0){\includegraphics[scale=0.5]{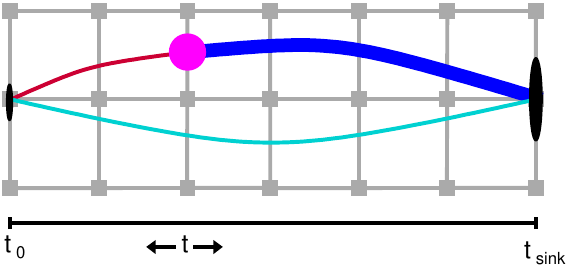}}
    \put(6,18){\textcolor{MyDarkRed}{$\bar c$}}
    \put(35,18){\textcolor{MyBlue}{$\bar b$}}
    \put(17,7.5){\textcolor{MyCyan}{$u/d$}}

    \put(63,0){\includegraphics[scale=0.5]{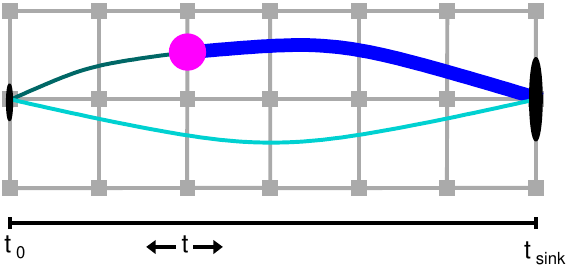}}
    \put(69,18){\textcolor{MyDarkGreen}{$\bar u$}}
    \put(98,18){\textcolor{MyBlue}{$\bar b$}}    
    \put(80,7.5){\textcolor{MyCyan}{$u/d$}}
  \end{picture}
  \caption{Sketch of the LQCD setup to calculate exclusive semileptonic $B\to D^* \ell\nu$ decays (left) and $B\to \pi\ell\nu$ decays (right).}
  \label{Fig.SketchExDeacay}
\end{figure}
    
Conventionally we parametrize semileptonic $B$ decays in terms of known kinematical terms ${\cal K}_{D^*}(q^2,m_\ell)$ and form factors ${\cal F}(q^2)$
\begin{align}
  \frac{d\Gamma(B\to D^*\ell\nu)}{dq^2}  = {\cal K}_{D^*}(q^2,m_\ell) \cdot |{\cal F}(q^2)|^2 \cdot |V_{cb}|^2.
\end{align} 
The form factors parametrize contributions due to the (nonperturbative) strong force and we use an OPE to identify short distance contributions. These short distance contribution are calculable using lattice QCD where the corresponding flavor changing currents are implemented as point-like operators. A sketch of the lattice setup for exclusive $B\to D^*\ell\nu$ decays is shown on the left hand side of Fig.~\ref{Fig.SketchExDeacay}. At the magenta dot the flavor changing vector ($V^\mu$) and axial ($A^\mu$) currents are inserted to calculate hadronic matrix elements and subsequently extract the (relativistic) form factors $V(q^2)$, $A_0(q^2)$, $A_1(q^2)$, and $A_2(q^2)$:

\begin{align}
  \langle D^*(k,\varepsilon_\nu) |{\cal V}^\mu | B(p)\rangle =& V(q^2) \frac{2i\varepsilon^{\mu\nu\rho\sigma}\varepsilon_\nu^* k_\rho p_\sigma}{M_{B}+M_{D}^*},%\\
\end{align}
\begin{align}
  \langle D^*(k,\varepsilon_\nu) |{\cal A}^\mu| B(p)\rangle=& {A_0(q^2)}\frac{2M_{D}^*\varepsilon^*\cdot q} {q^2}   q^\mu \notag\\
  &\quad + {A_1 (q^2)}(M_{B} + M_{D^*})\left[ \varepsilon^{*\mu}  - \frac{\varepsilon^*\cdot q}{q^2}   q^\mu \right] \notag\\
  &\quad -{A_2(q^2)} \frac{\varepsilon^*\cdot q}{M_{B}+M_{D}^*} \left[ k^\mu + p^\mu - \frac{M_{B}^2 -M_{D^*}^2}{q^2}q^\mu\right].
\end{align}
Since in a $b\to c$ transition a heavy bottom quark decays to a heavy charm quark, frequently the four form factors are expressed using the HQE convention where the momentum transfer $q^2$ is replaced by $w=v_{D^*}\cdot v_B$ and the four form factors are named $h_V(w)$, $h_{A_0}(w)$, $h_{A_1}(w)$, $h_{A_2}(w)$.

By now three lattice collaborations, Fermilab/MILC \cite{FermilabLattice:2021cdg}, JLQCD \cite{Aoki:2023qpa}, and HPQCD \cite{Harrison:2023dzh} have published form factor results for $B\to D^*\ell\nu$ at non-zero recoil. Fermilab/MILC and JLQCD restrict their lattice determinations to the range of high $q^2$ to keep cutoff effects well controlled. By first performing an extrapolation of the lattice data to physical quark masses and the continuum limit, they cover the full $q^2$ or $w$ range in a second step carrying out BGL $z$-expansion \cite{Boyd:1994tt,Boyd:1995sq}.  HPQCD follows a different strategy simulating heavy flavor masses ranging from charm-like to bottom-like masses. In a combined analysis HPQCD extrapolates their lattice data to the continuum with physical quark masses and performs the kinematical interpolation at the same time. An advantage of this strategy is that for heavy flavor masses below the bottom quark mass a larger, if not the entire phenomenologically allowed range of $q^2$ can be covered. The analysis is however more involved and direct comparisons/checks may be less straight forward.

In general these three form factor determinations show a reasonable level of consistency in particular for the range in $q^2$ directly covered by the individual lattice calculations. However, when considering form factors extrapolated over the full kinematically allowed range in $q^2$, tensions in the shape of the form factors show up warranting further scrutiny. Similarly when combining the form factor results with the binned experimental measurements by Belle \cite{Belle:2018ezy} and Belle II \cite{Belle-II:2023okj} tensions in the shape are present. Efforts are on-going to better understand the origin of these tensions, see e.g.~\cite{Bordone:2024weh}. Furthermore, additional groups are working on LQCD determinations of $B\to D^*\ell\nu$ form factors \cite{Bhattacharya:2020xyb, AnastasiaLattice2024}.

\section{Determination of \texorpdfstring{$|V_{ub}|$}{Vub}}
\label{Vub}
   
$|V_{ub}|$ is the least precisely known CKM matrix element. Although leptonic $B\to \tau \nu_\tau$ decays have been experimentally observed \cite{BaBar:2012nus,Belle:2015odw}, the uncertainties are too large to impact the determination of $|V_{ub}|$. Hence semileptonic decays are preferred but similarly to $|V_{cb}|$ these exhibit a long standing tension between determinations based on inclusive and exclusive decays. Here we report on recent updates concerning exclusive decays using LQCD to determine the nonperturbative input in terms of form factors. While for $|V_{cb}|$ the (narrow width) vector final state $D^*$ is the preferred channel for extracting the CKM matrix element, it is the pseudoscalar-to-pseudoscalar $B\to \pi\ell\nu$ decay in the case of $|V_{ub}|$. Conventionally we parametrize this process placing the $B$ meson at rest by
\begin{align}
  \frac{d\Gamma(B\to \pi\ell\nu)}{dq^2}  =& \frac{G_F^2 |V_{ub}|^2}{24 \pi^3} \,\frac{(q^2-m_\ell^2)^2\sqrt{E_\pi^2-M_\pi^2}}{ q^4M_{B}^2} \notag \\
  &\times \bigg[ \left(1+\frac{m_\ell^2}{2q^2}\right)M_{B}^2(E_\pi^2-M_\pi^2)|f_+(q^2)|^2
    +\,\frac{3m_\ell^2}{8q^2}(M_{B}^2-M_\pi^2)^2|f_0(q^2)|^2
    \bigg] \label{eq:B_semileptonic_rate}
\end{align}
and encode the nonperturbative input in terms of the two form factors $f_+$ and $f_0$. Again an OPE has been performed to identify the short distance contributions which we obtain from the lattice calculation by extracting the hadronic matrix element
\begin{align}
\langle \pi |V^\mu | B\rangle = f_+(q^2) \left( p^\mu_{B} + p^\mu_\pi - \frac{M^2_{B} - M^2_\pi}{q^2}q^\mu\right) + f_0(q^2)  \frac{M^2_{B} - M^2_\pi}{q^2}q^\mu.
\end{align}
A sketch of the lattice setup is shown on the right hand side in Fig.~\ref{Fig.SketchExDeacay}. Since pions are much lighter than $D^*$ mesons, $B\to\pi\ell\nu$ decays expand over a much larger kinematical range. So far all semileptonic form factor calculations for $B\to \pi\ell\nu$ on the lattice have only been performed at high $q^2$ and a kinematical $z$-extrapolation is performed to cover the entire range. Semileptonic form factors have been calculated by HPQCD \cite{Dalgic:2006dt}, RBC/UKQCD \cite{Flynn:2015mha}, Fermilab/MILC \cite{Lattice:2015tia}, and JLQCD \cite{Colquhoun:2022atw}. To combine the different lattice determinations, FLAG uses the continuum limit form factors from RBC/UKQCD, Fermilab/MILC, and JLQCD and extracts so called synthetic data points. Treating all calculations as statistically independent, a combined fit of these synthetic data points with the experimental measurements by BaBar \cite{delAmoSanchez:2010af,Lees:2012vv} and Belle \cite{Ha:2010rf,Sibidanov:2013rkk} using the BCL parametrization \cite{Bourrely:2008za} is performed.
The FLAG average value is
\begin{align}
  |V_{ub}^\text{excl}| = 3.64(16) \cdot 10^{-3},
\end{align}
where the error has been inflated following the PDG procedure for fits with poor $p$-value (large $\chi^2/\text{d.o.f.}$). Already the lattice form factors exhibit a small tension which may be caused by how the continuum limit of the form factors is taken.

This issue has been first pointed out in Ref.~\cite{Flynn:2023nhi} for semileptonic $B_s\to K\ell\nu$ decays, an alternative channel to determine the CKM matrix element $|V_{ub}|$. Form factors $f_+$ and $f_0$ describing semileptonic $B_s\to K\ell\nu$ decays over the full $q^2$ range have been obtained by HPQCD \cite{Bouchard:2014ypa}, RBC/UKQCD \cite{Flynn:2015mha}, and Fermilab/MILC \cite{FermilabLattice:2019ikx}. For several years the value at $q^2=0$ predicted by RBC/UKQCD and Fermilab/MILC has been in tension with the value predicted by HPQCD which is in turn consistent with analytic predictions \cite{Duplancic:2008tk,Wang:2012ab,Faustov:2013ima,Khodjamirian:2017fxg}. The lattice calculation for pseudoscalar final states typically proceeds by determining on the lattice the form factors $f_\parallel$ and $f_\perp$ which are directly accessible by hadronic matrix elements. Forming a linear combination of $f_\parallel$ and $f_\perp$ leads to the phenomenological form factors $f_+$ and $f_0$. As pointed out by RBC/UKQCD \cite{Flynn:2023nhi}, it is important to perform the chiral-continuum extrapolation using the phenomenological form factors $f_+$ and $f_0$ because only for phenomenological quantities pole masses entering the extrapolation formulae have a physical meaning. In the case of form factors describing $B_s\to K\ell\nu$ decays, Ref.~\cite{Flynn:2023nhi} demonstrates that using $f_+$ and $f_0$ in the chiral-continuum extrapolation (instead of $f_\parallel$ and $f_\perp)$ removes the tension. Furthermore, Flynn, Jüttner, and Tsang devised a new procedure based on Bayesian inference \cite{Flynn:2023qmi} to overcome issues related to truncating the $z$-expansion at too low order and find consistency with the dispersive matrix approach \cite{DiCarlo:2021dzg,Martinelli:2022tte}.

\section{Summary}\label{summary}
The determination of $|V_{cd}|$ seems to be in very good shape. Different determinations based on neutrino scattering, leptonic or semileptonic decays agree and the new Fermilab/MILC calculation using the full $q^2$ range in the semileptonic determination will help to reduce the uncertainty. Both inclusive and exclusive determinations of $|V_{cb}|$ have significantly progressed but the tension between both remains. Different inclusive determinations are consistent and the new method based on $q^2$ moments leads to further improvement. On the exclusive front we now have three independent determinations covering the full $q^2$ range. Although we observe some tension between the lattice form factors as well as w.r.t.~to the shape of the experimental data, having different data gives us a handle to further scrutinize these calculations and gain a better understanding.  $|V_{ub}|$ remains the CKM matrix element with the largest uncertainty. However, progress on the analysis of exclusive decay channels has been made and further work by different collaborations is ongoing. In addition new LQCD developments target the determination of inclusive processes on the lattice see e.g.~\cite{Hashimoto:2017wqo,Hansen:2017mnd,Bailas:2020qmv,Gambino:2020crt,Barone:2023tbl}.

% BibTeX or Biber users please use (the style is already called in the class, ensure that the "woc.bst" style is in your local directory)
\bibliography{B_meson.bib} % Replace "your_bib_file" with the actual name of your .bib file

\end{document}